\documentclass[11pt]{revtex4-1}

\usepackage{graphicx}

\begin{document}

\title{A complete cosmic scenario from inflation to late time acceleration: Non-equilibrium thermodynamics in the context of particle creation}

\author{Subenoy Chakraborty\footnote {schakraborty.math@gmail.com}}
\author{Subhajit Saha\footnote {subhajit1729@gmail.com}}
\affiliation{Department of Mathematics, Jadavpur University, Kolkata 700032, West Bengal, India.}

\begin{abstract}
The paper deals with the mechanism of particle creation in the framework of irreversible thermodynamics. The second order non-equilibrium thermodynamical prescription of Israel and Stewart has been presented with particle creation rate, treated as the dissipative effect. In the background of a flat FRW model, we assume the non-equilibrium thermodynamical process to be isentropic so that the entropy per particle does not change and consequently the dissipative pressure can be expressed linearly in terms of the particle creation rate. Here the dissipative pressure behaves as a dynamical variable having a non-linear inhomogeneous evolution equation and the entropy flow vector satisfies the second law of thermodynamics. Further, using the Friedmann equations and by proper choice of the particle creation rate as a function of the Hubble parameter, it is possible to show (separately) a transition from the inflationary phase to the radiation era and also from matter dominated era to late time acceleration. Also, in analogy to analytic continuation, it is possible to show a continuous cosmic evolution from inflation to late time acceleration by adjusting the parameters. It is found that in the de Sitter phase, the comoving entropy increases exponentially with time, keeping entropy per particle unchanged. Subsequently, the above cosmological scenarios has been described from field theoretic point of view by introducing a scalar field having self interacting potential. Finally, we make an attempt to show the cosmological phenomenon of particle creation as Hawking radiation, particularly during the inflationary era.

\end{abstract}

\maketitle
\section{INTRODUCTION}

It is generally speculated that non-equilibrium thermodynamical processes play a crucial role in the description of the physics of the early universe. In fact Schrodinger [1] initiated the microscopic description of the gravitationally induced particle production in an expanding universe. Then after a long gap, in late 1960's, Parker and others [2$-$4] discussed this issue based on quantum field theory in a curved spacetime. On the otherhand, from the thermodynamical viewpoint, Eckart [5] and Landau and Lifschitz [6] initiated the study of non-equilibrium phenomenon which is known as first order theory. But it suffers from serious drawbacks related to causality and stability. Subsequently, these drawbacks were removed by Muller [7], Israel [8], Israel and Stewart [9, 10], Pavon {\it et al.} [11], considering second order deviations from equilibrium. In this second order theory, dissipative phenomena like bulk and shear viscous pressure and heat flux are treated as dynamical variables having causal evolution equations such that thermal and viscous perturbations propagate at subluminal speeds.

In cosmology, homogeneous and isotropic model of the universe is commonly used in the literature, so bulk viscous pressure is the only dissipative phenomenon. This dissipative effect (bulk viscous pressure) may occur in FRW cosmology either due to coupling of different components of the cosmic substratum [12$-$16] or due to non conservation of (quantum) particle number [17$-$19]. The present work will mainly concentrate on this second aspect only and for simplicity of calculation, attention will be devoted to isentropic (or adiabatic) particle production [20, 21], i.e., production of perfect fluid particles. Although entropy per particle is constant (due to adiabatic character of the thermodynamical system), still there is entropy production due to enlargement of the phase space of the system (as the number of fluid particles increases). As a result, there will be a simple linear relationship between particle production rate and the viscous pressure, which will be used throughout the work.

Further, in the present work, Israel-Stewart type [9] entropy flow will be related to the cosmological particle production of isentropic nature and as a result, bulk viscous pressure will become a dynamical degree of freedom having causal evolution equation. Also, the drawbacks of bulk viscous inflation without particle production (as used by many authors) will be eliminated and it is possible to model a universe which starts in a de Sitter phase (having finite and stationary temperature and density) and subsequently evolves smoothly to standard FRW model. Also there will be attempts to exhibit late time acceleration within this causal, second order theory.

On the otherhand, the present effective imperfect fluid picture will be related to the dynamics of a scalar field from field theoretic point of view [22, 23] and it will allow us to express the growing entropy (due to paticle production) in a comoving volume in terms of scalar field variable. Finally, attempts will be made to associate this particle production with the phenomenon of Hawking radiation [24].

This paper is organized as follows: Sec. II deals with bulk viscous fluid in the perspective of non-equilibrium thermodynamics. In Sec. III, the particle production process is correlated with the second order non-equilibrium thermodynamics and also isentropic condition is used for simplicity. Sec. IV is devoted to the dynamics of FRW Universe with isentropic particle production on the basis of a second order expression for the entropy flow vector. The creation rate is regarded as a dynamical degree of freedom and by proper choice of this creation rate as a function of the Hubble parameter, both inflation and late time acceleration has been shown. A unified cosmic evolution has been described in Sec. V. A general expression for the change of entropy in a comoving volume as a fuction of the production rate and its simplified expression in de Sitter phase has been presented in Sec. VI. In Sec. VII, the mechanism of paticle production of the cosmic fluid has been shown to be equivalent to the dynamics of a scalar field from the field theoretic point of view. The cosmic particle creation is shown to be equivalent to the phenomenon of Hawking radiation (particularly during the inflationary era) in Sec. VIII. The paper ends with the summarization of the results obtained, in Sec. IX.
 
\section{NON-EQUILIBRIUM THERMODYNAMICS: BULK VISCOUS FLUID}

The energy-momentum tensor of a relativistic fluid having bulk viscosity as the only dissipative part is given by
\begin{equation}
T_{\mu \nu}=(\rho +p+\Pi)u_{\mu}u_{\nu}+(p+\Pi)g_{\mu \nu},
\end{equation}
where as usual $\rho$, $p$ and $\pi$ stands for the energy density, equilibrium pressure and bulk viscous pressure (connected with entropy production) and $u^{\mu}$ is the fluid 4-velocity. The conservation laws namely
\begin{equation} 
{T^{\mu \nu}}_{;\nu} =0~~~~and~~~~{N^\mu}_{;\mu} =0
\end{equation} 
has the explicit expressions
\begin{equation}
\dot{\rho}+\theta(\rho +p+\Pi)=0~~~~and~~~~\dot{n}+\theta n=0.
\end{equation}
Here $\theta ={u^\mu}_{;\mu}$ is the fluid expansion, $N^{\mu}=nu^{\mu}$ is the particle flow vector, $n$ is the particle number density and $\dot{n}=n_{,\alpha}u^{\alpha}$.

In second order non-equilibrium thermodynamics, the entropy flow vector ($S^{\alpha}$) according to Israel and Stewart, is given by [9, 10]
\begin{equation}
S^{\alpha}=sN^{\alpha}-\frac{\tau {\Pi}^2}{2\zeta T}u^{\alpha},
\end{equation}
where $s$ is the entropy per particle, $T$ is the fluid temperature, $\tau$ is the relaxation time and $\zeta$ is the coefficient of bulk viscosity.\\
Using the above conservation equations (3), the Gibb's relation
\begin{equation}
Tds=d\left(\frac{\rho}{n}\right)+pd\left(\frac{1}{n}\right)
\end{equation}
gives the variation of the entropy per particle as 
\begin{equation}
\dot{s}=-\frac{\Pi \theta}{nT}.
\end{equation}
Thus using Eqs. (3) and (6), we obtain the entropy production density from Eq. (4) as
\begin{equation}
{S^\alpha}_{;\alpha}=-\frac{\Pi}{T}\left[\theta +\frac{\tau}{\zeta}\dot{\Pi}+\frac{1}{2}\Pi\tau\left(\frac{\tau}{\zeta T}u^{\alpha}\right)_{;\alpha}\right].
\end{equation}
If we assume the following generalized linear relation
\begin{equation}
\Pi =-\zeta \left[\theta +\frac{\tau}{\zeta}\dot{\Pi}+\frac{\Pi T}{2}\left(\frac{\tau}{\zeta T}u^{\alpha}\right)_{;\alpha}\right],
\end{equation}
then from Eq. (7), we have,
\begin{equation}
{S^\alpha}_{;\alpha}=\frac{\Pi ^2}{\zeta T}\geq 0,
\end{equation}
i.e., second law of thermodynamics holds and the evolution equation for the viscous fluid pressure is obtained from Eq. (8) as
\begin{equation}
\Pi +\tau \dot{\Pi}=-\zeta \theta -\frac{1}{2}\Pi \tau \left[\theta +\frac{\dot{\tau}}{\tau}-\frac{\dot{\zeta}}{\zeta}-\frac{\dot{T}}{T}\right].
\end{equation}
Thus, in causal non-equilibrium thermodynamics, the dissipative pressure becomes a dynamical degree of freedom having evolution equation given by Eq. (10). Note that in the limit $\tau \rightarrow 0$, we have the first order Eckart theory [5] for which
\begin{equation}
S_E^{\alpha}=sN^{\alpha}~~~~and~~~~\Pi _E=-\zeta \theta.
\end{equation}
However, in between there is a truncated version of the non-equilibrium theory in which the evolution equation is a first order linear differential equation given by
\begin{equation}
\Pi +\tau \dot{\Pi}=-\zeta \theta,
\end{equation}
which is known as Maxwell-Cattaneo equation and is valid if the second term on the right hand side of Eq. (10) is negligible compared to the dissipative term '$-\zeta \theta$'. Thus truncated theory coincides with the full theory [23] if the term in the square bracket of Eq. (10) vanishes identically, i.e.,
\begin{equation}
\theta +\frac{\dot{\tau}}{\tau}-\frac{\dot{\zeta}}{\zeta}-\frac{\dot{T}}{T}=0.
\end{equation}
Now considering the particle number density '$n$' and the temperature '$T$' as the basic thermodynamical variables, the general form of the equations of state are
\begin{equation}
\rho =\rho(n,T)~~~~and~~~~p=p(n,T).
\end{equation}
Differentiating these equations of state and using the conservation equations (3) and the general thermodynamic relation
\begin{equation}
\frac{\partial \rho}{\partial n}=\frac{\rho +p}{n}-\frac{T}{n}\frac{\partial p}{\partial T},
\end{equation}
the evolution of temperature is given by
\begin{equation}
\frac{\dot{T}}{T}=-\theta \left[\frac{\Pi}{T(\frac{\partial \rho}{\partial T})}+\frac{\partial p}{\partial \rho}\right].
\end{equation}
The above equation shows that in case of viscous fluid, the temperature variation depends on the dissipative factor $\Pi$. As $\Pi$ is normally chosen to be negative, so the first term on the right hand side will counteract the second term and consequently close to the equilibrium, i.e., for $|\Pi|<p$, the temperature of the thermodynamical system will decrease less rapidly than in the perfect fluid case. Hence one can say that the bulk viscous pressure has a tendency to 'reheat' the thermodynamical system.

Similarly, the evolution of the thermodynamical pressure is given by
\begin{equation}
\dot{p}=-\theta \left[(\rho +p)c_s^{2}+\Pi \frac{\partial p}{\partial \rho}\right],
\end{equation}
where the adiabatic sound velocity $c_s$ is given by [23]
\begin{equation}
c_s^{2}=\left(\frac{\partial p}{\partial \rho}\right)_{ad.}=\frac{n}{\rho +p}\frac{\partial p}{\partial n}+\frac{T}{\rho +p}\frac{{\left(\frac{\partial p}{\partial T}\right)}^2}{\left(\frac{\partial \rho}{\partial T}\right)}.
\end{equation}
Further, the velocity of propagation of the viscous pulses is obtained as [25, 26]
\begin{equation}
c_b^{2}=\frac{\zeta}{(\rho +p)\tau}
\end{equation}
and hence the sound in a viscous medium propagates with velocity $v$ ($<1$), where [26]
\begin{equation}
v^2=c_s^{2}+c_b^{2}.
\end{equation} 
It should be noted that due to causality, we have $c_b^{2}\leq 1-c_s^{2}$. Now using the time variation of temperature and the expressions for the speed of sound and viscous pulses the evolution Eq. (10) now becomes [23]
\begin{equation}
\Pi +\tau \dot{\Pi}=-\rho \theta \tau \left[\gamma c_b^{2}+\frac{\Pi}{2\rho}\left\lbrace 2+c_s^{2}+\frac{\partial p}{\partial \rho}\right\rbrace+\frac{{\Pi}^2}{2\gamma {\rho}^2}\left\lbrace 1+\frac{\partial p}{\partial \rho}+\frac{(\rho +p)}{T\left(\frac{\partial \rho}{\partial T}\right)}\right\rbrace \right]+\frac{\tau \Pi}{2}\frac{{(c_b^{2})}^{.}}{c_b^{2}}
\end{equation}
where $p=(\gamma -1)\rho$ is the barotropic equation of state. It should be noted that the above nonlinear evolution equation holds for arbitrary (or even arbitrary time varying) equations of state.

\section{PARTICLE PRODUCTION AND NON-EQUILIBRIUM THERMODYNAMICS}

We shall now consider an open thermodynamical system where the number of fluid particles is not preserved. So the second conservation equation in (3) is to be modified as $(N^a_{;a}\neq 0)$ [27, 28]
\begin{equation}
N^a_{;a}\equiv \dot{n}+\theta n=n\Gamma,
\end{equation}
where $\Gamma$ stands for the rate of change of the number of particles $(N=na^3)$ in a comoving volume $a^3$. Clearly, positivity of $\Gamma$ indicates the creation of particles while $\Gamma <0$ stands for particle annihilation. It is to be noted that a nonvanishing $\Gamma$ will produce an effective bulk pressure [19, 27, 29$-$33] on the thermodynamic fluid and one can use the non-equilibrium thermodynamics. The above statement can simply be demonstrated in case of 'isentropic' particle production as follows:

From Gibb's equation (i.e., Eq. (5)), using the (modified) conservation equations, we have
\begin{equation}
nT\dot{s}=-\Pi \theta -\Gamma (\rho +p).
\end{equation}
But due to isentropic (i.e., adiabatic) process, the equilibrium entropy per particle does not change (as it does in dissipative process (in Sec. II)), i.e., $\dot{s}=0$ and we obtain [27, 28]
\begin{equation}
\Pi =-\frac{\Gamma}{\theta}(\rho +p).
\end{equation}
Hence the bulk viscous pressure is entirely determined by the particle production rate. So we may say that a dissipative fluid is equivalent to a perfect fluid having varying particle number. Note that although $\dot{s}=0$, still there is entropy production due to the enlargement of the phase space of the system.

At first we consider the entropy flow vector in Eckart theory, i.e., (suffix '$E$' stands for Eckart theory)
\begin{equation}
S_E^{\alpha}=nsu^{\alpha}.
\end{equation}
Then,
\begin{equation}
{S_{E}^{\alpha}}_{;\alpha}={\Gamma _{E}}\left(ns-\frac{\rho +p}{T}\right)-\frac{\Pi _E \theta}{T},\\
~~i.e.,~ {TS_E^{\alpha}}_{;\alpha}=-n\mu \Gamma _E -\Pi _E \theta,
\end{equation}
where $\mu =\frac{\rho +p}{n}-Ts$ is the chemical potential. So we have,
\begin{equation}
{S_E^{\alpha}}_{;\alpha}=\frac{\Pi _E^{2}}{\zeta T}\geq 0,
\end{equation}
where the effective bulk pressure $\Pi _E$ satisfies the quadratic equation
\begin{equation}
\Pi _E^{2}+\zeta \theta \Pi _E+n\zeta \mu \Gamma _E=0.
\end{equation}
Further, if we consider the isentropic process, then
\begin{equation}
{S_E^{\alpha}}_{;\alpha}=-\frac{nsT}{\rho +p}\frac{\Pi _E\theta}{T}.
\end{equation}
So if we define
\begin{equation}
\Pi _E=-\zeta \left(\frac{nsT}{\rho +p}\right)\theta ~~with~\zeta >0,
\end{equation}
then
\begin{equation}
{S_E^{\alpha}}_{;\alpha}=\frac{\Pi _E^{2}}{\zeta T}\geq 0~~and~~
\Gamma _E=\zeta \frac{nsT}{(\rho +p)^2}{\theta}^2>0.
\end{equation}
So, only positive $\Gamma _E$ is compatible with the second law of thermodynamics.

On the otherhand, if we consider Eq. (4) as the entropy flow vector, then proceeding as above (in Sec II), we have,
\begin{equation}
T{S^{\alpha}}_{;\alpha}=-n\mu \Gamma -\Pi \left[\theta +\frac{\tau \dot{\Pi}}{\zeta}+\frac{1}{2}\Pi T{\left(\frac{\tau}{\zeta T}u^{\alpha}\right)}_{;\alpha}\right].
\end{equation}
Thus, choosing the generalized ansatz as
\begin{equation}
\theta +\frac{\tau}{\zeta}\dot{\Pi}+\frac{1}{2}\Pi T{\left(\frac{\tau}{\zeta T}u^{\alpha}\right)}_{;\alpha}+\frac{\mu n\Gamma}{\Pi}=-\frac{\Pi}{\zeta},
\end{equation}
we have, ${S^\alpha}_{;\alpha}=\frac{{\Pi}^2}{\zeta T}\geq 0$ and the effective viscous pressure $\Pi$ is determined by a nonlinear inhomogeneous differential equation
\begin{equation}
{\Pi}^2+\tau \Pi \dot{\Pi}+\frac{1}{2}\zeta {\Pi}^2 T{\left(\frac{\tau}{\zeta T}u^{\alpha}\right)}_{;\alpha}+\zeta \Pi \theta =-\zeta \mu n\Gamma .
\end{equation}
Note that in absence of chemical potential, the above evolution equation reduces to a linear first order differential equation. The chemical potential $\mu$ may act as an effective symmetry-breaking parameter in relativistic field theories.

The essential physical difference between the noncausal and the causal theory is the introduction of the time of relaxation (in the later one). If a nonvanishing $\Gamma$ is responsible for the occurence of an effective viscous pressure, then in causal theory there is a finite relaxation time during which $\Pi$ decays to zero after $\Gamma$ has been switched off. Also, due to this nonvanishing particle production rate, the evolution of the temperature will be modified from the expression in Eq. (16). Proceeding as before and using Eq. (22) instead of the second equation of (3), we obtain [23]
\begin{equation}
\frac{\dot{T}}{T}=-\theta \left[\frac{\frac{\partial p}{\partial T}}{\frac{\partial \rho}{\partial T}}+\frac{\Pi}{T\frac{\partial \rho}{\partial T}}\right]+\Gamma \left[\frac{\frac{\partial p}{\partial T}}{\frac{\partial \rho}{\partial T}}-\frac{(\rho +p)}{T\frac{\partial \rho}{\partial T}}\right],
\end{equation}
or equivalently using Eq. (23),
\begin{equation}
\frac{\dot{T}}{T}=-(\theta -\Gamma)\frac{\frac{\partial p}{\partial T}}{\frac{\partial \rho}{\partial T}}+\frac{n\dot{s}}{\frac{\partial \rho}{\partial T}}.
\end{equation}
Thus, for finite non-zero particle production rate, the variation of temperature is not only characterized by the effective viscous pressure $\Pi$ but also there is an additional direct coupling of the production rate $\Gamma$. 

Further, in this second order theory, if we again impose the isentropic condition (Eq. (24)), then
\begin{equation}
{S^{\alpha}}_{;\alpha}=-\Pi \left[\frac{\Pi}{2}\left(\frac{\tau}{\zeta T}u^{\alpha}\right)_{;\alpha}+\frac{\tau}{\zeta T}\dot{\Pi}-\frac{ns\Gamma}{\Pi}\right].
\end{equation}
Also, due to this 'isentropic' particle production, the relevant thermodynamical variables evolve as [23]
\begin{equation}
\dot{\rho}=-(\theta -\Gamma)(\rho +p)~,~~\dot{p}=-c_s^{2}(\theta -\Gamma)(\rho +p)~,~~\frac{\dot{n}}{n}=-(\theta -\Gamma)~,~~\frac{\dot{T}}{T}=-(\theta -\Gamma)\frac{\partial p}{\partial \rho}.
\end{equation}
Now using (38) in the expression (37), we have 
\begin{equation}
T{S^{\alpha}}_{;\alpha}=nsT\Gamma -\frac{\tau}{\zeta}\left[(\rho +p)\frac{\Gamma}{\theta}\right]^{2}\left[\left\lbrace \frac{{\left(\frac{\Gamma}{\theta}\right)}^{.}}{\frac{\Gamma}{\theta}}-(1+c_s^{2})(\theta -\Gamma)\right\rbrace+\frac{1}{2}\left\lbrace \theta +\frac{{(\frac{\tau}{\zeta})}^{.}}{\frac{\tau}{\zeta}}-\frac{\dot{T}}{T}\right\rbrace \right],
\end{equation}
where in the last equation, we have used the isentropic condition (24). Now introducing the speed for viscous pulses (see Eq. (19)), Eq. (39) takes the form
\begin{equation}
T{S^{\alpha}}_{;\alpha}=(\rho +p)\frac{\Gamma}{\theta}\left[\frac{nsT}{(\rho +p)}\theta -(\rho +p)\left(\frac{\Gamma}{\theta}\right)\left(\frac{\tau}{2\zeta}\right) \left\lbrace \Gamma +2\frac{{\left(\frac{\Gamma}{\theta}\right)}^{.}}{\left(\frac{\Gamma}{\theta}\right)}-(\theta -\Gamma)\left(c_s^{2}-\frac{\partial p}{\partial \rho}\right)-\frac{{\left(c_b^{2}\right)}^{.}}{c_b^{2}}\right\rbrace \right].
\end{equation}
Now for the generalized linear relation namely,
\begin{equation}
(\rho +p)\frac{\Gamma}{\theta}=\zeta \left[\frac{nsT}{(\rho +p)}\theta -(\rho +p)\left(\frac{\Gamma}{\theta}\right)\left(\frac{\tau}{2\zeta}\right)\left\lbrace \Gamma +2\frac{{\left(\frac{\Gamma}{\theta}\right)}^{.}}{\left(\frac{\Gamma}{\theta}\right)}-(\theta -\Gamma)\left(c_s^{2}-\frac{\partial p}{\partial \rho}\right)-\frac{{\left(c_b^{2}\right)}^{.}}{c_b^{2}}\right\rbrace \right]
\end{equation}
with $\zeta >0$, we have the generalized second law of thermodynamics for the entropy production density
\begin{equation}
{S^{\alpha}}_{;\alpha}=\frac{\left[(\rho +p)\frac{\Gamma}{\theta}\right]^{2}}{\zeta T}\geq 0.
\end{equation}
Eq. (41) can be recast as the dynamical equation for $\frac{\Gamma}{\theta}$ [23], i.e.,
\begin{equation}
\left(\frac{\Gamma}{\theta}\right)+\tau {\left(\frac{\Gamma}{\theta}\right)^{.}}=\theta \tau \left[\frac{nsT}{\rho +p}c_b^{2}-\frac{1}{2}{\left(\frac{\Gamma}{\theta}\right)}^{2}\left(1+c_s^{2}-\frac{\partial p}{\partial \rho}\right)+\frac{1}{2}\left(\frac{\Gamma}{\theta}\right)\left(c_s^{2}-\frac{\partial p}{\partial \rho}\right)+\frac{1}{2}\left(\frac{\Gamma}{\theta}\right)\frac{{\left(c_b^{2}\right)}^{.}}{c_b^{2}}\right].
\end{equation}
The above equation can be converted to the evolution equation for $\Pi$ using the isentropic condition (24). Further, as $\tau \rightarrow 0$ we get back to the Eckart's first order relation.

\section{PARTICLE CREATION IN FRW UNIVERSE: INFLATION AND LATE TIME ACCELERATION}

As the particle number is not conserved, so we shall consider an open model of the universe. Here, for simplicity, we consider spatially flat FRW model of the universe having line element
\begin{equation}
ds^{2}=-dt^2+a^{2}(t)\left[dr^2+r^2d{\Omega_2}^2\right].
\end{equation}
Now the Einstein field equations for the cosmic fluid having energy-momentum tensor given by Eq. (1) are
\begin{equation}
\kappa \rho =3H^2~~and~~\dot{H}=-\frac{\kappa}{2}(\rho +p+\Pi),
\end{equation}
where $\kappa$ is the Einstein's gravitational constant and $H=\frac{\theta}{3}$ is the Hubble parameter. It should be noted that the cosmic fluid may be considered as a perfect fluid and the dissipative term $\Pi$ is the effective bulk viscous pressure (see the previous section) due to particle production, i.e., the cosmic substratum is not a conventional dissipative fluid, rather a perfect fluid with varying particle number.

In the context of cosmology, as the substratum is a perfect fluid, so we shall focus on isentropic particle production, i.e., a process having constant entropy per particle. However, there will be entropy production due to the enlargement of the phase space of the system since the universe is expanding and also the number of fluid particles increases. So the effective bulk pressure does not characterize a conventional non-equilibrium, rather a state having equilibrium properties as well (but not the equilibrium era with $\Gamma =0$). Hence compared to the second order non-equilibrium thermodynamics, the relaxation time $\tau$ has been interpreted in the context of cosmology as a relaxation from states with $\dot{s}=0$, $\Gamma>0$ to states with $\dot{s}=0=\Gamma$. Here $\tau$ is of the order of Hubble time and the process is not related to interactions in which the mean free collision time plays a role. Now using the Einstein field equations (45) into the isentropic condition (24), we have
\begin{equation}
\frac{\Gamma}{\theta}=1+\frac{2}{3\gamma}\frac{\dot{H}}{H^2}.
\end{equation}

\subsection*{Case I: First order Eckart theory}

In the first order non-equilibrium thermodynamics, the rate of change of particle number for isentropic process is given by Eq. (31), i.e.,
\begin{equation}
\Gamma _E=\zeta \frac{nsT}{(\rho +p)^2}\theta ^2.
\end{equation}
If the chemical potential is absent then using the first Einstein field equation, it turns out that $\Gamma _E$ is constant (say $\Gamma _0$), i.e.,
\begin{equation}
\Gamma _0=\frac{3\zeta \kappa}{\gamma}.
\end{equation}
So, in Eckart theory, with isentropic process and vanishing chemical potential always leads to constant rate of particle creation. Now solving the differential equation (46) in $H$, we obtain,
\begin{equation}
H=\frac{\Gamma _0}{3}+\left({\frac{a}{a_0}}\right)^{-\frac{3\gamma}{2}},
\end{equation}
and
\begin{equation}
a=a_1\left[e^{\frac{\gamma {\Gamma _0}}{2}(t-t_0)}-1\right]^{\frac{2}{3\gamma}},
\end{equation}
with $a_1=a_0\left(\frac{3}{\Gamma _0}\right)^{\frac{2}{3\gamma}}$, $a_0$ and $t_0$ are integration constants. Thus the universe starts with a big bang singularity at $t=t_0$ and then evolves exponentially.

\subsection*{Case II: Second order Israel-Stewart theory}

In this case if we use Eq. (46) in Eq. (43), then the evolution equation for $H$ becomes a second order nonlinear differential equation as follows: [23]
\begin{eqnarray}
\tau H \big[\frac{\ddot{H}}{H} &-& \frac{{\dot{H}}^2}{\gamma H^2}\left(1+c_s^{2}+\frac{\partial p}{\partial \rho}\right)+3\dot{H}\left\lbrace 1-\frac{1}{2}\left(\frac{\partial p}{\partial \rho}-c_s^{2}\right)\right\rbrace -\frac{9\gamma H^2}{2}\left(c_b^{2}\frac{nTs}{\rho +p}-\frac{1}{2}\right)-\frac{1}{2}\frac{\left(c_b^{2}\right)^{.}}{Hc_b^{2}}\left(\dot{H}+\frac{3}{2}\gamma H^2\right)\big] \nonumber \\
&+& \dot{H}+\frac{3\gamma}{2}H^2=0.
\end{eqnarray}
If we consider the early de Sitter phase where $\dot{H}=0$ then from Eq. (46), $\Gamma =3H=constant$, irrespective of the equation of state of the cosmic fluid and consequently all the relevant thermodynamical parameters namely $n$, $T$, $\rho$ and $p$ are constants (see the evolution Eqs. (38)). Further, in the de Sitter phase, if the chemical potential is neglected then from the evolution equation (51), we obtain,
\begin{equation}
\tau H=\frac{1}{3\left(c_b^{2}-\frac{1}{2}\right)}
\end{equation}
which also does not depend on the equation of state of the fluid. From the above equation we see that the relaxation time is of the order of Hubble time and compared to conventional causal theory, $\tau$ is enlarged. This 'frozen in' of the non-equilibrium description can be considered as a necessary condition for successful inflation. Also, from the above equation, we must have $c_b^{2}>\frac{1}{2}$, which is compatible with the general causality restriction $c_b^{2}\leq 1-c_s^{2}\leq \frac{2}{3}$. Further, due to the evolution of the universe, the above de Sitter solution is no longer a stable one and for $\Gamma \ll 3H$, we have the usual cosmological solution without particle production.

In the radiation dominated era ($p=\frac{1}{3}\rho$), the evolution equation (51) takes the form (with vanishing chemical potential)
\begin{equation}
\tau H \left[\frac{\ddot{H}}{H}-\frac{5}{4}\frac{{\dot{H}}^2}{H^2}+3\dot{H}-6H^2 \left(c_b^{2}-\frac{1}{2}\right)\right]+\dot{H}+2H^2=0.
\end{equation}
Due to its complicated form we cannot solve it exactly and hence from Eq. (46), we cannot determine the ratio $\frac{\Gamma}{\theta}$ which characterizes the dynamics. 

However, for inflationary regime, the ratio $\frac{\Gamma}{\theta}$ is constant while particle production is strongly suppressed (i.e., $\frac{\Gamma}{\theta}\ll 1$) in entering the radiation era, so phenomenologically to describe this early period of inflation the simplest choice is the following linear relationship, i.e., $\frac{\Gamma}{\theta}\propto H$. So we write [23]
\begin{equation}
\Gamma =3\beta \frac{H^2}{H_r},
\end{equation} 
where $\beta$ is the constant of proportionality and $H_r$ is the Hubble parameter at some fixed epoch $t_r$ (with $a_r=a(t_r)$). For this choice of ansatz, Eq. (46) can be solved to obtain (with arbitrary equation of state parameter $\gamma$)
\begin{equation}
H=\frac{H_r}{\beta +(1-\beta)\left(\frac{a}{a_r}\right)^{\frac{3\gamma}{2}}}.
\end{equation}
Thus as $a\rightarrow 0$, $H\rightarrow {\beta}^{-1}H_r=constant$, indicating an exponential expansion ($\ddot{a}>0$) in the inflationary era while for $a\gg a_r$, $H\propto a^{-\frac{3\gamma}{2}}$ represents the standard FRW cosmology ($\ddot{a}<0$). Suppose we identify $a_r$ as some intermediate value of '$a$' where $\ddot{a}=0$ (i.e., the transition epoch from the de Sitter stage to the standard radiation era). Then we have $\dot{H}_r=-H_r^{2}$ and from Eq. (46),
\begin{equation}
\beta =1-\frac{2}{3\gamma}.
\end{equation}
Hence for relativistic matter (i.e., for radiation $\gamma =\frac{4}{3}$),
\begin{equation}
\beta =\frac{1}{2}~~~~and~~~~H=\frac{2H_r}{1+\left(\frac{a}{a_r}\right)^2}.
\end{equation}
which on integration gives
\begin{equation}
t=t_r +\frac{1}{4H_r}\left[ln\left(\frac{a}{a_r}\right)^2 +\left(\frac{a}{a_r}\right)^2 -1 \right].
\end{equation}
Thus during the early evolution of the universe we have the following limiting situations (for relativistic matter, i.e., $\gamma = \frac{4}{3}$):
\begin{eqnarray}
a &\sim & a_r e^{2H_r t}~~~for~a\ll a_r~~~(Early~inflationary~era) \\
a &\sim & a_r t^{\frac{1}{2}}~~~~~~~for~a\gg a_r~~~(Standard~cosmological~regime).
\end{eqnarray}
Also, in the standard cosmological regime, the rate of particle production $\Gamma$ decreases as inverse square law, i.e., $\Gamma \sim t^{-2}$.



Further, if we now assume that at the end of the matter dominated era $\Gamma$ becomes insignificant again for standard particles, {\it i.e., } baryons become almost conserved but the production of dark sector particles becomes gradually significant. So we again describe the transition phenomenologically as
\begin{equation}
\frac{\Gamma}{\theta}\propto \frac{1}{H^2},~i.e.,~ \Gamma =3\delta \frac{H_f}{H},
\end{equation}
where $\delta$ is the proportionality constant and $H_f$ is the value of the Hubble parameter at some instant $t_f$ where $\ddot{a}=0$ in course of the transition from deceleration ($\ddot{a}<0$) to acceleration ($\ddot{a}>0$) (i.e., in quintessence era). Using Eq. (61) in Eq. (46) and integrating, we obtain
\begin{equation}
H^2=\delta H_f +\left(\frac{a}{a_f}\right)^{-3\gamma}.
\end{equation}
So for the late time evolutionary scenario, we have the limiting situations as (for arbitrary $\gamma$):
\begin{eqnarray}
H &\sim & a^{-\frac{3\gamma}{2}}~~~~~when~a\ll a_f~~~(Standard~FRW~evolution) \\
H &\sim & \sqrt{\delta H_f}~~~when~a\gg a_f~~~(Late~time~accelerated~expansion).
\end{eqnarray}
Also, one can determine the expression for the relaxation time $\tau$ in terms of the deceleration parameter $q$ $\left(=-\left(1+\frac{\dot{H}}{H^2}\right)\right)$ as
\begin{equation}
\tau =\frac{2(1-q)}{H \left[6\left(q+2c_b^{2}\right)-(1+q)(1+q-3\gamma)\right]}.
\end{equation}

\section{A UNIFIED COSMIC EVOLUTION}

Though in the last section, we have mentioned that the particle creation rates are chosen phenomenologically, however there are some thermodynamical arguments behind the choices (particularly those which describe the inflation and the late time acceleration).

In the very early universe, (starting from a regular vacuum) most of the particle creation effectively takes place and from thermodynamical point of view \cite{Lima1},\\\\
$\bullet$ At the beginning of the expansion, there should be a maximal entropy production rate (i.e., maximal particle creation rate) so that the universe evolves from non-equilibrium thermodynamical state to equilibrium era with the expansion of the Universe.\\
$\bullet$ There should be a regular (true) vacuum for radiation initially, i.e., $\rho \rightarrow 0$ as $a \rightarrow 0$.\\
$\bullet$ Also, one should have $\Gamma >H$ in the very early Universe so that the created radiation behaves as a thermalized heat bath and subsequently the creation rate should fall slower than the expansion rate and particle creation becomes dynamically insignificant.\\\\
Now, according to Gunzig {\it et al.} \cite{Gunzig1}, the simplest choice satisfying the above requirements is that the particle creation rate is proportional to the energy density, {\it i.e., $\Gamma =3\beta \frac{H^2}{H_r}$}, where $H_r$ has the dimension of Hubble parameter (i.e., reciprocal of time). Thus $\beta$ has the dimension of $M^{-1}T^{-1}$.

For the intermediate deceleration phase, the simplest natural choice is $\Gamma \propto H$. It should be noted that this choice of $\Gamma$ does not satisfy the third thermodynamical requirement at the early universe (mentioned above).

Further, for the late time cosmic acceleration phase, the thermodynamical requirements of the early epoch are modified as:\\\\
$\bullet$ There should be minimum entropy production rate at the beginning of the late time accelerated expansion and the universe again becomes thermodynamically non-equilibrium.\\
$\bullet$ The late time false vacuum should have $\rho \rightarrow 0$ as $a \rightarrow \infty$.\\
$\bullet$ The creation rate should be faster than the expansion rate.\\\\
It can be shown that another simple choice of $\Gamma$, namely $\Gamma \propto \frac{1}{H}$ will satisfy these requirements.

We have chosen the time instants $t=t_r$ and $t=t_f$ (in the previous section) as the time of transitions from acceleration to deceleration (in the early phase) and again from deceleration to acceleration (in the late era). Thus we shall denote the particle creation rates as\\\\
$\bullet$ $\Gamma =3\beta \frac{H^2}{H_r}$~~~for $t \leq t_r$~~(called phase I)\\
$\bullet$ $\Gamma =3\Gamma _0 H$~~~for $t_r \leq t \leq t_f$~~(called phase II)\\
$\bullet$ $\Gamma =3\delta \frac{H_f}{H}$~~~~for $t \geq t_f$~~(called phase III)\\\\
The cosmological solutions with relevant thermodynamical parameters in these phases are given below.\\\\
{\bf Phase I:}
\begin{eqnarray}
H &=& \frac{H_r}{\beta +(1-\beta)\left(\frac{a}{a_r}\right)^{\frac{3\gamma}{2}}} 
\nonumber \\ \nonumber \\ \nonumber
\rho &=& \rho _r \left[\beta +(1-\beta){\left(\frac{a}{a_r}\right)}^{\frac{3\gamma}{2}}\right]^{-2} 
\nonumber \\ \nonumber \\ \nonumber 
n &=& n_r \left[\beta +(1-\beta){\left(\frac{a}{a_r}\right)}^{\frac{3\gamma}{2}}\right]^{-\frac{2}{\gamma}}
\nonumber \\ \nonumber \\ \nonumber 
T &=& T_r \left[\beta +(1-\beta){\left(\frac{a}{a_r}\right)}^{\frac{3\gamma}{2}}\right]^{-\frac{2(\gamma -1)}{\gamma}},
\nonumber
\end{eqnarray} 
where $a_r$, $\rho _r$ ($=\frac{3H_{r}^{2}}{\kappa}$), $n_r$ and $T_r$ are the values of the scale factor, energy density, number density and temperature at the early transition time $t_r$. The Hubble parameter can be expressed in terms of the cosmic time $t$ as
\begin{center}
$H=H_r \left[\beta \left\lbrace LambertW\left(\frac{1-\beta }{\beta} exp\left\lbrace \frac{2(1-\beta)+3\gamma H_r(t-t_r)}{2\beta}\right\rbrace \right)+1 \right\rbrace \right]^{-1}$,
\end{center}
where the function $LambertW(x)$ is defined by $LambertW(x)e^{LambertW(x)}=x$. Note that for $\gamma =\frac{4}{3}$ (radiation epoch), we have $\rho \sim T^4$, the usual black body radiation.\\\\
{\bf Phase II:} 
\begin{eqnarray}
H &=& H_r \left(\frac{a}{a_r}\right)^{-\frac{3\gamma (1-\Gamma _0)}{2}}
\nonumber \\ \nonumber \\ \nonumber
\rho &=& \rho _r \left(\frac{a}{a_r}\right)^{3\gamma (1-\Gamma _0)}
\nonumber \\ \nonumber \\ \nonumber
n &=& n_r \left(\frac{a}{a_r}\right)^{3(1-\Gamma _0)}
\nonumber \\ \nonumber \\ \nonumber
T &=& T_r \left(\frac{a}{a_r}\right)^{3(\gamma -1)(1-\Gamma _0)}.
\nonumber 
\end{eqnarray}
One can also write the Hubble parameter as
\begin{center}
$H=H_r \left[1+\frac{3\gamma}{2} H_r \left(1-\Gamma _0\right) \left(t-t_r\right)\right]^{-1}$,
\end{center}
in terms of the cosmic time $t$.\\\\
{\bf Phase III:}
\begin{eqnarray}
H &=& \left[\delta H_f +\left(\frac{a}{a_f}\right)^{-3\gamma}\right]^{\frac{1}{2}}
\nonumber \\ \nonumber \\ \nonumber
\rho &=& \rho _f \left[\frac{1}{H_{f}^{2}}\left\lbrace {\delta H_f}+\left(\frac{a}{a_f}\right)^{-3\gamma} \right\rbrace \right]
\nonumber \\ \nonumber \\ \nonumber
n &=& n_f \left[\frac{1}{H_{f}^{2}}\left\lbrace {\delta H_f}+\left(\frac{a}{a_f}\right)^{-3\gamma} \right\rbrace \right]^{\frac{1}{\gamma}}
\nonumber \\ \nonumber \\ \nonumber
T &=& T_f \left[\frac{1}{H_{f}^{2}}\left\lbrace {\delta H_f}+\left(\frac{a}{a_f}\right)^{-3\gamma} \right\rbrace \right]^{\frac{\gamma -1}{\gamma}},
\nonumber
\end{eqnarray}
where $a_f$, $\rho _f$ ($=\frac{3H_{f}^{2}}{\kappa}$), $n_f$ and $T_f$ are the values of the scale factor, energy density, number density and temperature at the late transition time $t_f$. Here also, the Hubble parameter can be expressed explicitly in terms of the cosmic time $t$ as
\begin{center}
$H=\sqrt{\delta H_f} coth\left\lbrace \frac{3\gamma}{2}\sqrt{\beta}H_f (t-t_1)\right\rbrace$,
\end{center}
where $t_1$ is the constant of integration.

We shall now examine whether the cosmic evolution across the different phases are smooth or not. For smoothness of the particle creation rate at the transition epochs $t=t_r$ and $t=t_f$, we must have
\begin{equation}
\Gamma _0=\beta ~~~~and~~~~\delta =\beta H_f.
\end{equation}
It is clear from the first two solution sets that Hubble parameter, scale factor and the physical parameters namely energy density, number density and temperature are continuous across $t=t_r$. For the continuity across $t=t_f$, we must have
\begin{equation} \label{x}
(1-\beta)H_{f}^{2}=1~,~~~~\frac{3\gamma}{2}(1-\beta)(t_f-t_r)=\frac{1}{H_f}-\frac{1}{H_r}.
\end{equation}
\begin{equation} \label{y}
\left(\frac{a_r}{a_f}\right)^{\frac{3\gamma (1-\beta)}{2}}=\frac{H_f}{H_r}=\frac{1}{H_r \sqrt{1-\beta}}.
\end{equation}
and
\begin{equation} \label{z}
sinh(\mu _f)=\sqrt{\delta H_f}=\sqrt{\beta} H_f~,~~~~cosh(\mu _f)=H_f, \nonumber \\
\end{equation}
where 
\begin{center}
$\mu _f=\frac{3\gamma}{2}\sqrt{\beta}H_f (t_f-t_1)$.
\end{center}

Thus $t_1$ is determined from relations (\ref{z}) while $t_f$, $a_f$ and $H_f$ are evaluated from Eqs. (\ref{x}) and (\ref{y}). Also, the interrelation between the values of the Hubble parameter and the scale factor at the two transition epochs are shown in Eq. (\ref{y}).

The above solutions in the three phases are smooth across the transition epochs $t=t_r$ and $t=t_f$. The solutions contain four arbitrary parameters namely $a_r$, $H_r$, $t_r$ and $\beta$. To have some idea about these free parameters, we consider some observational constraints namely\\\\
a) The e-folding number of the inflation $N \geq 60$,\\
b) The reheating temperature $T>1 MeV$,\\
c) The relative energy density of dark energy $\Omega _d \simeq 0.7$.\\\\
As $a_r$ is the value of the scale factor at the end of inflation, so $a_r$ is constrained by the e-folding number. The reheating temperature restricts $T_r$, the temperature at the transition epoch $t_r$ and thereby $H_r$ is restricted by the reheating temperature. Also, the time of reheating gives an estimate of the transition time $t_r$. At phase III, if we consider matter as a combination of dark matter (in the form of dust) and dark energy, then either by using the field equations or from the solution in phase III, we can calculate the deceleration parameter to give
\begin{center}
$1+\Omega _d \omega _d=\gamma (1-\beta)$.
\end{center}
Thus from the observed value of $\Omega _d$ ($\simeq 0.7$) and the equation of state parameter $w_d$ ($\simeq -1.04$) \cite{Wang1}, we can estimate $\beta$. Hence in principle, we can estimate all the free parameters from the observational constraints.

Further, in Figs. 1(A)$-$1(C), we have been able to demonstrate the smooth evolution of the Hubble parameter, the particle number density and the temperature respectively, from the early inflationary era to late time acceleration.  Considering $\beta =\frac{1}{3}$, $\gamma =\frac{3}{2}$, $H_r=3$, $a_r=\frac{1}{3}$, $n_r=2$ and $T_r=\frac{5}{2}$, we have evaluated the late time transition values as $H_f=1.2247$, $a_f=0.6057$, $n_f=0.6057$ and $T_f=1.3758$, using the above continuity conditions. The deceleration parameter $q$ for our model has also been plotted in Fig. 2. 

Further, in order to compare our results with observational data, let us expand the scale factor in a power series about the present time (to describe the late time cosmic acceleration) as \cite{Guimaraes1}
\begin{equation}
\frac{a(t)}{a(t_0)}= 1+ H_0 (t-t_0)-\frac{1}{2!} q_0 H_{0}^2 (t-t_0)^2+ \frac{1}{3!} j_0 H_{0}^3 (t-t_0)^3+\frac{1}{4!} s_0  H_{0}^4 (t-t_0)^4+ O[(t-t_0)^5],
\end{equation}
where $H_0$, $q_0$, $j_0$ and $s_0$ are the values of Hubble, deceleration, jerk and snap parameters at the present epoch $t_0$. Then the deceleration parameter can be expressed in terms of the redshift parameter $z$ as
\begin{equation}
q(z)= q_0+ (-q_0-2 q_{0}^2+ j_0)z+\frac{1}{2} (2q_0+ 8q_{0}^2+ 8q_{0}^3-7q_0j_0-4j_0-s_0)z^2+ O(z^3).
\end{equation}
Note that the deceleration parameter $q$ can as well be expressed in terms of the scale factor $a$ by using the substitution $z=\left(\frac{1}{a}-1 \right)$ in the above equation.

\begin{figure}
\includegraphics[width=8cm, height=8cm]{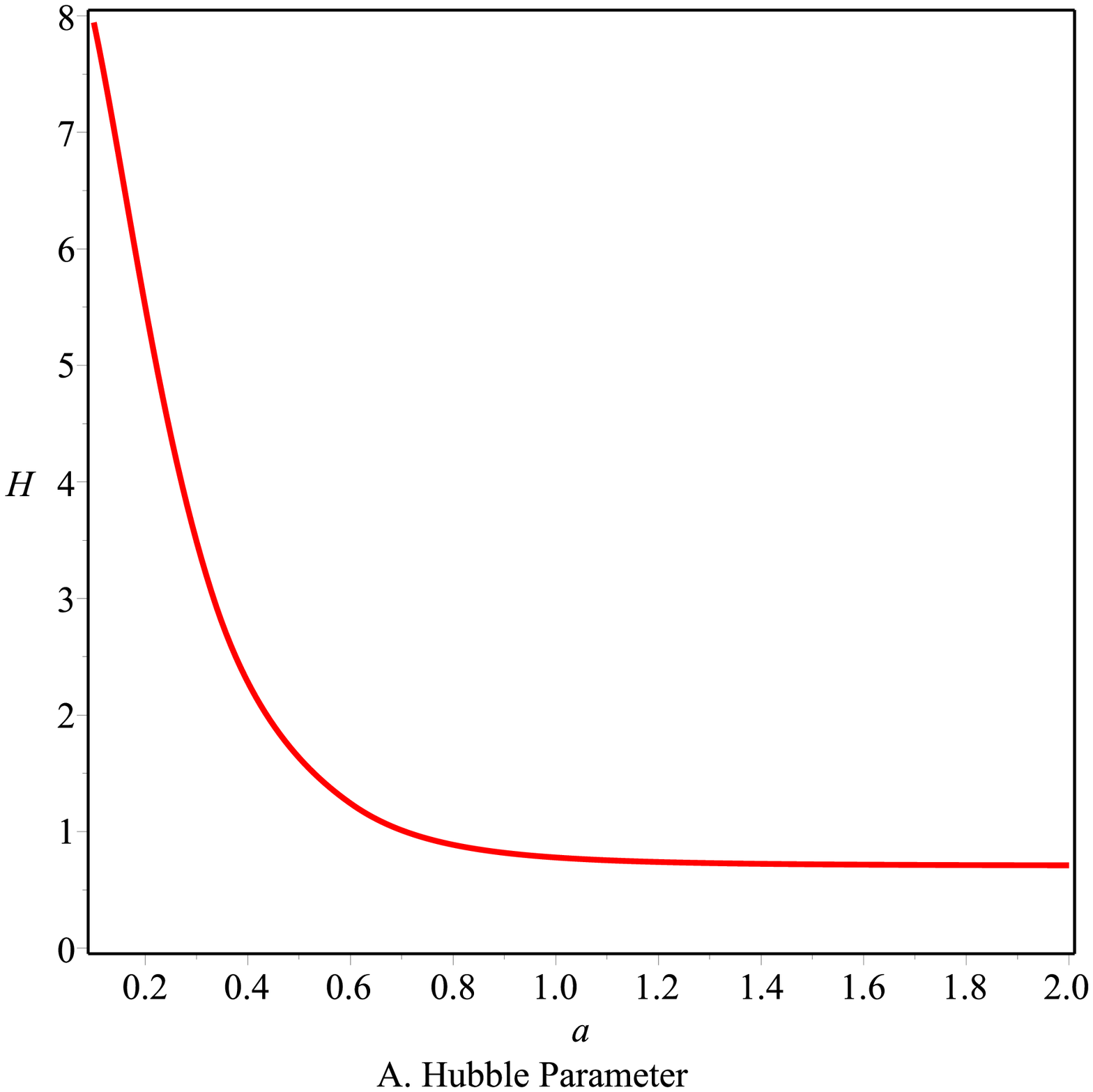}
\end{figure}

\begin{figure}
\begin{minipage}{0.4\textwidth}
\includegraphics[width=1.0\linewidth]{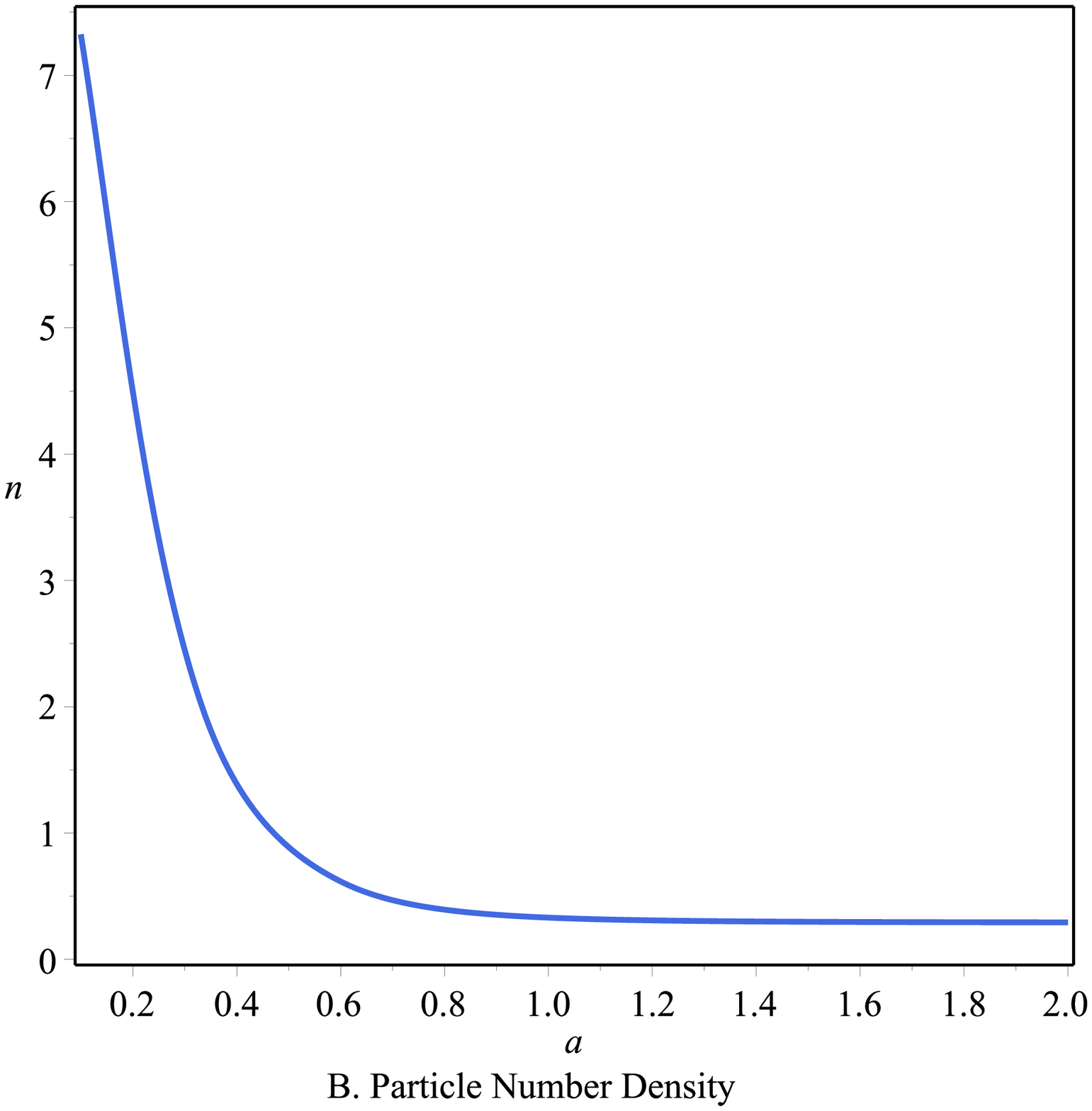}
\end{minipage}
\begin{minipage}{0.4\textwidth}
\includegraphics[width=1.0\linewidth]{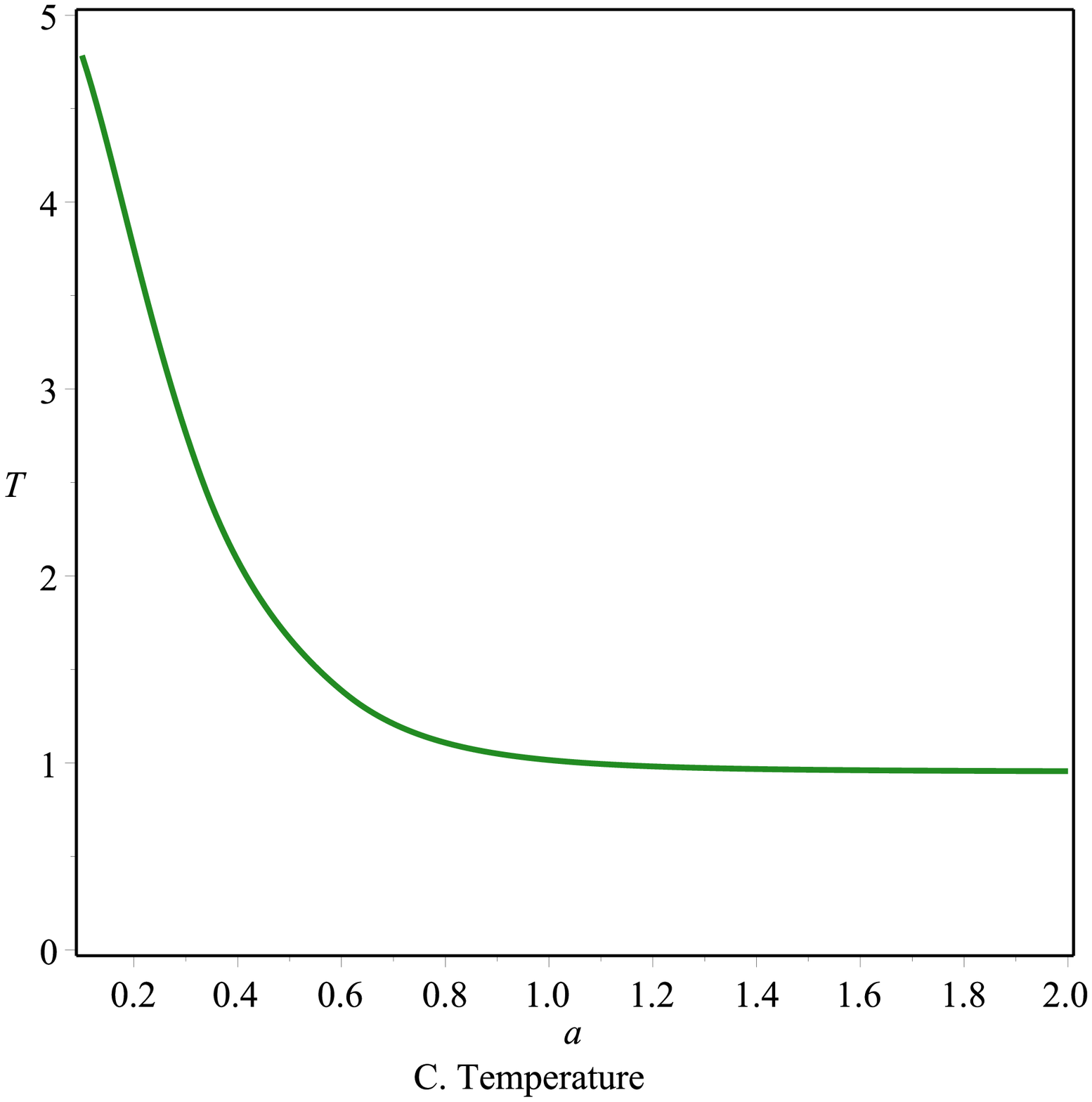}
\end{minipage}
\caption{The figures A, B and C respectively show the evolution of the Hubble parameter, the particle number density and the temperature against the scale factor $a$.}
\end{figure}

\begin{figure}
\includegraphics[width=8cm, height=8cm]{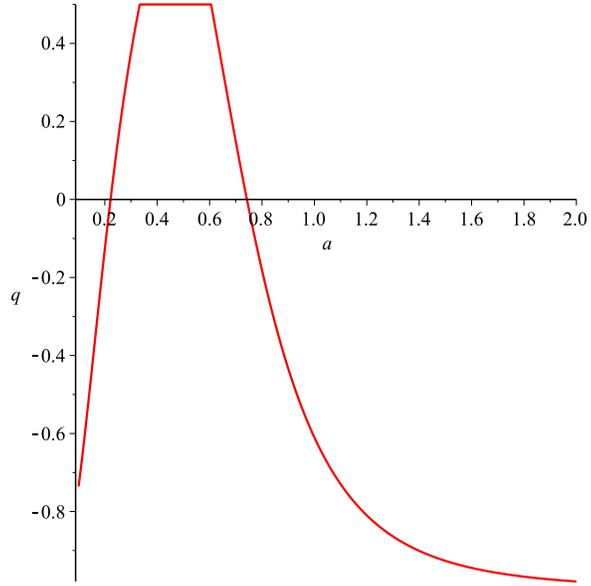}
\caption{The figure shows the evolution of the deceleration parameter against the scale factor $a$.}
\end{figure}

\begin{figure}
\includegraphics[width=8cm, height=8cm]{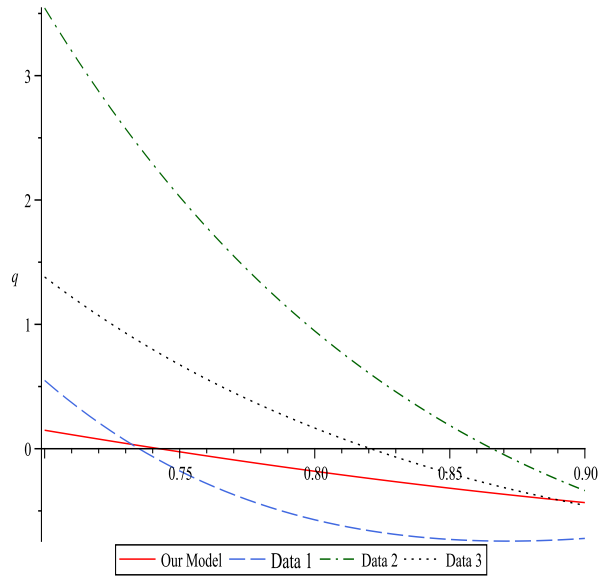}
\caption{The figure shows the comparison of the late time evolution of the deceleration parameter for our model with that for data sets 1, 2 and 3 against the scale factor $a$.}
\end{figure}

Now, we shall attempt to compare our model with three observed data sets, namely 192 SNe Ia and 69 GRBs
with CPL Parametrization (Data 1) and Linear Parametrization (Data 2) \cite{Wang1}, and Union 2+BAO+OHD+GRBs Data (Data 3) \cite{Xu1}, which have been presented in Table I.

\begin{center} {\bf Table I}: Observed data sets \end{center} 
\begin{center}
\begin{tabular}{|c|c|c|c|}
\hline Data & $q_0$ & $j_0$ & $s_0$\\
\hline \hline 1. 192 SNe Ia+GRBs (CPL Parametrization) & $-0.90$ & $3.93$ & $-25.52$\\
\hline 2. 192 SNe Ia+GRBs (Linear Parametrization) & $-0.75$ & $2.21$ & $-12.25$\\
\hline 3. Union 2+BAO+OHD+GRBs & $-0.386$ & $-4.925$ & $-26.404$\\
\hline
\end{tabular}
\end{center}
In Fig. 3, we have plotted the deceleration parameter for our model and for the above three observed data sets about the present time. From the figure, we see that our estimated $q$ matches clearly with the observed data of CPL parametrization than the other two observed data sets.

\section{ENTROPY PRODUCTION}

If the thermal process of the cosmic fluid is not isentropic then the evolution of the entropy per particle is given by Eq. (23). From the Einstein field equations (45), the bulk pressure has the expression [23]
\begin{equation}
\Pi =-\rho \left[\gamma +\frac{2}{3} \frac{\dot{H}}{H^2}\right]
\end{equation}
and we have (using the conservation equation (3))
\begin{equation}
\frac{\dot{\rho}}{\rho}=2\frac{\dot{H}}{H}.
\end{equation}
Now substituting $\Pi$ from Eq. (72) and using Eq. (73), the temperature evolution equation (16) can be written as 
\begin{equation}
\frac{\dot{T}}{T}=(\gamma u-v)\left(-\frac{\dot{n}}{n}\right)+u\frac{\dot{\rho}}{\rho},
\end{equation}
with $u=\frac{\rho}{T\frac{\partial \rho}{\partial T}}$, $v=\frac{\frac{\partial p}{\partial T}}{\frac{\partial \rho}{\partial T}}$. If $u$, $v$ and $\gamma$ are assumed to be constants then integration of Eq. (74) gives [23]
\begin{equation}
T=T_i \left(\frac{\rho}{\rho _i}\right)^u \left(\frac{a^3 N_i}{{a_i}^3 N}\right)^{\gamma u-v},
\end{equation}
where subscript '$i$' stands for the value of the variables at some initial time. Also using Eqs. (72) and (73), the rate of change of the entropy per particle (given by Eq. (23)) can be written as
\begin{equation}
nT\dot{s}=(\rho +p)(3H-\Gamma)+\dot{\rho}=\frac{d}{dt}\left(\frac{\rho a^{3\gamma}}{N^\gamma}\right)\frac{N^\gamma}{a^{3\gamma}},
\end{equation}
or using Eq. (75) we can write
\begin{equation}
\dot{s}=\frac{1}{n_i T_i}\left(\frac{\rho _i}{\rho}\right)^u \left(\frac{{a_i}^3 N}{a^3 N_i}\right)^{\gamma u-v-1} \left(\frac{N}{a^3}\right)^\gamma \left(\frac{\rho a^{3\gamma}}{N^\gamma}\right).
\end{equation}
For simplicity, if we choose the inflationary phase having $\rho =\rho _i$, $H=H_i$ and $\gamma =\frac{4}{3}$ then equation (79) simplifies to
\begin{equation}
\dot{s}=\frac{4\rho _i}{n_i T_i}\left[H_i -\frac{\Gamma}{3}\right]
\end{equation}
which on integration gives
\begin{equation}
s=s_i +\frac{4\rho _i H_i}{n_i T_i}(t-t_i)-\frac{4\rho _i}{3n_i T_i}\int _{t_i}^t {\Gamma dt}.
\end{equation}
Thus the change of entropy in a moving volume has the explicit expression (in inflationary phase)
\begin{equation}
S_{cv}=nsa^3=\left[\frac{4\rho _i}{n_i T_i}\left(H_i(t-t_i)-\frac{1}{3}\int _{t_i}^t {\Gamma dt}\right)+s_i \right]N_iexp \left\lbrace \int _{t_i}^t {\Gamma dt} \right\rbrace.
\end{equation}
Note that due to Eq. (22) the above equation is no longer characterized by change of '$s$' alone. Further, if $\Gamma =\Gamma _i$, a constant, then Eq. (80) becomes [23]
\begin{equation}
S_{cv}=N_i \left[\frac{4\rho _i (t-t_i)}{n_i T_i}\left(H_i-\frac{1}{3}\Gamma _i \right)+s_i \right]exp \left\lbrace {\Gamma _i (t-t_i)}\right\rbrace.
\end{equation}
Hence for $\Gamma \neq 0$, in the de Sitter phase, the comoving entropy increases exponentially with time. Also the bulk viscous pressure $\Pi$ ($<0$) in the Einstein equations (45) not only gives a repulsive gravity effect, it is even connected to an increase in the number of fluid particles rather than an increase in the entropy per particle.

\section{FIELD THEORETIC DESCRIPTION: SCALAR FIELD MODEL}

In the earlier sections, we have shown both early epoch of inflationary phase as well as a period of late time acceleration by considering matter creation models without introducing any DE component (or a cosmological constant). But in order to describe the cosmological scenario from the view point of field theory, it is desirable to address the whole dynamical process as the evolution of a scalar field $\phi$ having self interacting potential $V(\phi)$ or in other words we shall show how the effective imperfect fluid dynamics (discussed in this work) can be viewed as the evolution of a minimally coupled scalar field. So the energy density and thermodynamic pressure of the cosmic fluid are given by
\begin{equation}
\rho =\frac{1}{2}{\dot{\phi}}^2 +V(\phi)~~~~and~~~~p_{eff}=p+\Pi=\frac{1}{2}{\dot{\phi}}^2 -V(\phi)
\end{equation}
and hence
\begin{equation}
{\dot{\phi}}^2=(\rho +p+\Pi)~~~~and~~~~V(\phi)=\frac{1}{2}(\rho -p-\Pi).
\end{equation}
Now using the isentropic condition (24), the particle creation rate as Eq. (54) (for the early epoch) and the expression for the Hubble parameter from Eq. (57), we obtain
\begin{equation}
\phi={\phi}_0+\frac{2}{\sqrt{\kappa}}sinh^{-1}\left(\frac{a}{a_r}\right)~,
~~~i.e.,~a={a_r}sinh \left[\frac{\sqrt{\kappa}}{2}\left(\phi -{\phi}_0 \right)\right]
\end{equation}
and 
\begin{equation}
V(\phi)=\frac{4H_{r}^{2}}{\kappa}sech^{2}\psi\left[1+2sech^{2}\psi \right]~,~~~\psi=\frac{\sqrt{\kappa}}{2}\left(\phi -{\phi}_0 \right).
\end{equation}
Suppose at the inflationary epoch ($t=t_I$), $a_I$ be the value of the scale factor and $V_I$ is the constant value of the potential and the scalar field has the value
\begin{equation}
{\phi}_I={\phi}_0+\frac{2}{\sqrt{\kappa}}sinh^{-1}\left(\frac{a_I}{a_r}\right).
\end{equation}
(Note that the scalar field has always a value greater than ${\phi}_0$). Now during inflationary scenario, from the point of view of slow-roll approximation, the density fluctuations are of the form ${\delta}_H\sim \frac{H^2}{{\phi}^2}\sim 10^{-5}$ [22, 33]. In the present model [22]
\begin{equation}
\frac{H^2}{{\dot{\phi}}^2}=2\pi G\left(1+\frac{a_{r}^{2}}{a^2}\right)~,~~~i.e.,~\frac{H^2}{{\dot{\phi}}^2}|_{a=a_I}=2\pi G \left\lbrace 1+\left(\frac{a_r}{a_I}\right)^2 \right\rbrace.
\end{equation}
Thus ${\delta}_H\sim 2\pi G \left(1+\frac{a_{r}^{2}}{a_{I}^{2}}\right)$, i.e.,
\begin{equation}
\left(\frac{a_r}{a_I}\right)^2\sim \frac{{\delta}_H}{2\pi G}-1\sim \frac{{\delta}_H}{2\pi {l_{pl}}^{2}},~ i.e.,~\frac{a_r}{a_I}\sim \left(\sqrt{\frac{{\delta}_H}{2\pi l_{pl}}}\right)^{-1},
\end{equation}
where $l_{pl}\simeq \sqrt{G}$ is the Planck length in units ${\hbar}=c=1$. So the ratio of the scale factors at the end of inflation to the start of inflation is proportional to the inverse of the Planck length or equivalently the comoving volume increases during inflation by the order of ${l_{pl}}^{-3}$. The particle creation rate and the comoving entropy (entropy in comoving volume) can be expressed in terms of the scalar field as
\begin{equation}
\Gamma =6H_r sech^{4}\psi
\end{equation}
and
\begin{equation}
S_{cv}={N_i}\left(\frac{tanh^{3}\phi}{tanh^3{\phi}_0}\right)\left[\frac{4\rho _i}{n_i T_i}\left\lbrace H_i(t-t_i)-log\left(\frac{tanh\phi}{tanh{\phi}_0}\right)\right\rbrace +s_i \right].
\end{equation}
This effective scalar field entropy of a comoving volume may be attributed to the universe, independently of the specific matter model.

Subsequently, in the matter dominated era, when the universe expands in a power law fashion ($a\sim t^l$), the particle creation rate is proportional to the Hubble parameter and we have
\begin{equation}
\phi={\phi}_0+\left(\sqrt{\frac{8l}{3\kappa}}\right) ln\left(\frac{t}{t_0}\right),
\end{equation}
\begin{equation}
V(\phi)=\frac{3l^2}{2\kappa t_{0}^{2}}\left\lbrace 2-(1-{\Gamma}_0)\gamma \right\rbrace exp\left[-\left(\sqrt{\frac{3\kappa}{2l}}\right)(\phi-{\phi}_0)\right],
\end{equation}
\begin{equation}
\Gamma=\frac{\Gamma _0}{t}
\end{equation}
and
\begin{equation}
S_{cv}={N_i}{\left(\frac{t}{t_i}\right)}^{\Gamma _0}\left[\frac{4\rho _i}{n_i T_i}\left\lbrace H_i(t-t_i)-\frac{\Gamma _0}{3}ln\left(\frac{t}{t_i}\right)\right\rbrace +s_i \right].
\end{equation}

Lastly, to describe the late time acceleration, we have used (in the previous section) the particle production rate as (see Eq. (61)) $\Gamma =3\delta \frac{H_f}{H}$ and then the Hubble parameter and the scale factor are given by the Eqs. (62) and (71). Hence as before we obtain
\begin{equation}
\phi={\phi}_0+\frac{2}{3\sqrt{\gamma \kappa}}\left[cosechT-cothT \right]~,~~~T=\frac{3\gamma}{2}\sqrt{\delta H_f}(t-t_0),
\end{equation}
\begin{equation} 
V(\phi)=3\delta\frac{H_f}{2\gamma}\left[2coth^{2}T-\gamma cosech^{2}T \right],
\end{equation}
\begin{equation}
\Gamma =3\sqrt{\delta H_f}tanhT
\end{equation}
and 
\begin{equation}
S_{cv}={N_i}\left(\frac{coshT}{cosh{T_i}}\right)^{\frac{2}{\gamma}}\left[\frac{4\rho _i}{n_i T_i}\left\lbrace H_i(t-t_i)-\frac{2}{\gamma}log\left(\frac{coshT}{cosh{T_i}}\right)\right\rbrace +s_i \right].
\end{equation}
One should note that in this section '$t_i$' stands for some initial reference time and subscript '$i$' in any variable indicates the value of the variable at the initial reference time $t_i$. Thus the use of the fluid or the scalar field picture and conversion between them gives a comprehensive overall picture of the evolution dynamics of the universe. Also it may lead to establish the correlation between the fluid cosmology and particle physics motivated study of the early universe.

\section{Particle creation as a phenomenon of Hawking radiation}

We have seen that due to particle creation, there is a negative pressure term in the Einstein field equations and it is considered as effective bulk viscous pressure. By proper choice of the particle creation rate, it has been possible to show that the universe underwent an inflationary era at the early stages of its evolution as well as it expereiences late time acceleration . We shall now try to correlate this phenomenon of particle creation as Hawking radiation from FRW space-time.

At first we shall discuss the Hawking radiation from FRW space-time bounded by the apparent horizon which characterizes the local properties of the space-time and its position is determined by the relation [38]
\begin{equation}
(h^{ab}\partial_a R\partial_b R)|_{R=R_A}=0,
\end{equation}
where $h_{ab}$ is the metric of the 2-space $x^a$=($t$, $r$) by writing the FRW metric as
\begin{equation}
ds^2=h_{ab}dx^a dx^b +R^2 d\Omega_{2}^{2}
\end{equation}
with $R=ar$ as the area radius and $h_{ab}=diag \left(-1,\frac{a^2}{\sqrt{1-k r^2}}\right)$. Now the solution of Eq. (99) gives
\begin{equation}
R_A=\frac{1}{\sqrt{H^2+\frac{k}{a^2}}}.
\end{equation}
The surface gravity of the apparent horizon is
\begin{equation}
\kappa=\frac{1}{R_A}\left(1-\frac{\dot{R}_A}{2HR_A}\right)
\end{equation}
and hence the Hawking temperature is given by [36]
\begin{equation}
T=\frac{\hbar \kappa}{2\pi k_B}=\frac{\hbar}{2\pi k_B}\left(\frac{1}{R_A}\right)\left(1-\frac{\dot{R}_A}{2HR_A}\right).
\end{equation}
In the early inflationary era, the scale factor '$a$' expands exponentially so that the Hubble parameter is a constant ($H_0$). If $H_0$ is such that ${H_0}^2\gg \frac{1}{a^2}$ (this is possible near the Planck size of the universe, i.e., $a\simeq l_p=10^{-35}m$ and so $\frac{c^2}{a^2}\simeq 10^{87}$, while at that scale $H_0=\frac{\dot{a}}{a}\simeq 10^{45} sec^{-1}$), then $R_A=\frac{1}{H}$ and as $H$ is constant, $\dot{R}_A=0$ and hence [39] $T\simeq \frac{\hbar H}{2\pi k_B}\sim 10^{32} K$.

In BH evaporation, at the start of the Hawking radiation, the size of the BH is enormous and evaporation process is very weak (for a BH of mass $\sim$ $M_0$, $T\sim 10^{-7} K$) and subsequently as the size of the BH decreases, the process (as well as the temperature rise) gradually becomes faster and faster. Finally, at the end phase, when the size of the BH is of Planck size, then thermal spectrum of Hawking radiation should be replaced by some quantum gravity effect. On the other hand, the interpretation of Hawking radiation in the inflationary era of the FRW model of the universe is just the reverse one. Here, at the beginning, the size of the universe is of the Planck size, so quantum gravity effects are dominant and gradually with the expansion of the universe, Hawking radiation comes into the picture and inflation starts. Subsequently, as the size of the universe gradually increases, the temperature gradually decreases and inflationary process will automatically turn off.

Also the present process of Hawking radiation is the inverse of BH evaporation. For BH, the created particles escape outside the event horizon towards asymptotic infinity, while in the FRW space-time, particles created near the apparent horizon will move inside the horizon. Further, due to isotropy of the FRW space-time, the radiation is isotropic from all directions. Thus in case of BH there is a loss of energy while for the FRW space-time, the universe gains energy which can be expressed by the Stephen-Boltzmann radiation law as [40]
\begin{equation}
P=\frac{dQ}{dt}=\sigma A_H T^4,
\end{equation}  
where $\sigma=\frac{\pi ^2 {k_B}^2}{60 {\hbar}^3 c^2}$ is the Stephen-Boltzmann constant and $T=\frac{\hbar H}{2\pi k_B}$ is the Hawking temperature. So from the first law of thermodynamics, {\it i.e., }
\begin{equation}
\frac{dQ}{dt}=\frac{d}{dt}(\rho V)+p\frac{dV}{dt}.
\end{equation}
We have [38]
\begin{equation}
\dot{\rho}+3H(\rho +p)=3\sigma HT^4.
\end{equation}
Using the first conservation equation (3) and the isentropic relation (24), we have 
\begin{equation} \label{xx}
\Gamma =\frac{\sigma}{\gamma}\frac{T^4}{H}
\end{equation}
or using the expression for the Hawking temperature (i.e., $T=\frac{\hbar H}{2\pi k_B}$), we have $\Gamma \propto H^3$, i.e., $\frac{\Gamma}{3\theta}=\lambda H^2$. Then from Eq. (46), we have
\begin{equation}
H^{-2}=\lambda +\left(\frac{a}{a_0}\right)^{3\gamma}.
\end{equation}
Hence $H\sim \frac{1}{\sqrt{\lambda}}$ for $a\ll a_0$ and $H\sim a^{-\frac{3\gamma}{2}}$ for $a\gg a_0$.

Thus we have as usual, the exponential expansion in the early phase followed by the evolution in standard cosmology. One should note that although the choice (54) for particle production rate gives the required evolution of the universe but it is not described by the Hawking radiation as according to Stephen-Boltzmann radiation law, $T\propto H^{\frac{3}{4}}$ (not $T\propto H$). Further, in phase II (the intermediate stage of evolution), $\Gamma \propto H$, so from Eq. (\ref{xx}), $T \propto H^\frac{1}{2}$. Hence in this case also, there is Hawking type radiation but not according to Stephen-Boltzmann radiation law while at the late time evolution (i.e., hase III), $T$ turns out to be constant and we may conclude that the particle production rate at the late time do not have an analogy to Hawking type radiation.


\section{Summary of the results}

This paper deals with non-equilibrium thermodynamics in the context of cosmology having perfect fluid as cosmic substratum. The dissipative phenomenon occurs due to particle creation mechanism and behaves as a bulk viscous pressure. We have employed the second order non-equilibrium thermodynamical prescription of Israel and Stewart, so that the dissipative pressure behaves as a dynamical variable having a non-linear inhomogeneous evolution equation and the entropy flow vector satisfies the second law of thermodynamics. For simplicity, we have assumed the thermodynamical process to be adiabatic, i.e., the entropy per particle remains constant and as a result the dissipative pressure is related linearly to the particle creation rate. Due to a complicated form of the evolution equation of the dissipative pressure, it is not possible to solve it, rather phenomenologically we have assumed the dissipative pressure (or the particle creation rate) as a function of the Hubble parameter. By proper choice of the functional form and considering a flat FRW model (open themrodynamical system), we are able to show the evolution of the Universe from inflationary phase to radiation era, standard cosmological evolution in matter dominated era and then a transition to late time acceleration without introduction of any dark energy. A thermodynamical argument in support of the choices of the particle creation rate has been given in Sec. V. In the three phases, complete cosmological solution as well as the relevant thermodynamical parameters have been evaluated. By proper choice of the parameters involved, it is possible to show a smooth transition of the cosmic parameters as well as the thermodynamic parameters across $t_r$ and $t_f$. Although the entropy per particle is constant still we have determined the comoving entropy as an integral of the particle creation rate. It is found that in the de Sitter phase, the comoving entropy increases exponentially with time.

Further, to have a field theoretic description, we have modeled the effective imperfect fluid dynamics as the evolution of a minimally coupled scalar field. The conversion between the fluid and the scalar field picture gives a comprehensive overall evolution of the universe. Also, this interrelation can act as the correlation between the fluid cosmology and the particle physics motivated study of the early universe. Lastly, we have tried to interpret the particle production as a phenomenon of Hawking radiation. We have found that during inflationary phase, due to particle production, the temperature has the form of black body radiation and hence we have concluded that the particle production in the inflationary era can be viewed as Hawking radiation. On the other hand, corresponding to late time acceleration, the temperature turns out to be constant and hence it cannot be considered as Hawking radiation. 

Finally, we conclude that in the framework of particle creation mechanism, Einstein gravity shows a unified cosmic picture from the eraly inflationary era to the present day late time acceleration without introducing any concept of dark energy. Therefore, particle creation mechanism naturally exhibit the observed late time acceleration without the introduction of dark energy or a modification of the Einstein gravity.

\begin{acknowledgments}

The authors are thankful to IUCAA, Pune, India for their warm hospitality and research facilities as the work was there during a visit. Also SC acknowledges the UGC-DRS Programme in the Department of Mathematics, Jadavpur University. The author SS is thankful to UGC-BSR Programme of Jadavpur University for awarding JRF.

\end{acknowledgments}

\end{document}